\documentclass[final,3p,times]{elsarticle}

\usepackage{wrapfig}

\usepackage{graphics}
\usepackage{graphicx}
\usepackage{epsfig}
\usepackage{amssymb}
\usepackage{amsthm}
\usepackage{amsmath}

\usepackage{caption}
\usepackage[labelfont=bf]{caption}
\usepackage[labelsep=space]{caption}
\usepackage[nodots, compress]{numcompress}
\biboptions{sort&compress}

\usepackage{xcolor}
\usepackage{hyperref}
\hypersetup{
colorlinks,
citecolor=[violet],
linkcolor=[red],
urlcolor=[blue]
}

\journal{Materials Today Energy}

\begin{document}

\begin{frontmatter}

\title{MgF$_2$ as an effective additive for improving ionic conductivity of ceramic solid electrolytes}

\author[1]{Pengfei Zhou}
\author[1]{Kaitong Sun}
\author[1]{Shunping Ji}
\author[1]{Zirui Zhao}
\author[1]{Ying Fu}
\author[1]{Junchao Xia}
\author[2]{Si Wu}
\author[1]{Yinghao Zhu}
\author[1]{Kwun Nam Hui}
\author[1]{Hai-Feng Li\corref{cor1}}
\cortext[cor1]{Corresponding author}
\ead{haifengli@um.edu.mo}

\affiliation[1]{organization={Joint Key Laboratory of the Ministry of Education, Institute of Applied Physics and Materials Engineering, University of Macau},
            addressline={Avenida da Universidade, Taipa},
            city={Macao SAR},
            postcode={999078},
            country={China}}
\affiliation[2]{organization={School of Physical Science and Technology, Ningbo University},
            city={Ningbo},
            postcode={315211},
            country={China}}

\begin{abstract}
As typical solid-state electrolytes (SSEs), {Na}$_{1+x}${Zr}$_2${Si}$_{x}${P}$_{3-x}${O}$_{12}$ NASICONs provide an ideal platform for solid-state batteries (SSBs) that display higher safety and accommodate higher energy densities. The critical points for achieving SSBs with higher efficiencies are to improve essentially the ionic conductivity and to reduce largely the interfacial resistance between SSEs and cathode materials, which would necessitate extremely high level of craftsmanship and high-pressure equipment. An alternative to higher-performance and lower-cost SSBs is additive manufacturing. Here, we report on an effective additive, MgF$_2$, which was used in synthesizing NASICONs, resulting in SSEs with fewer defects and higher performance. With an addition of mere 1 wt$\%$ MgF$_2$ additive, the total room-temperature ionic conductivity of the NASICON electrolyte reaches up to 2.03 mS cm$^{-1}$, improved up to $\sim$ 181.3$\%$, with an activation energy of 0.277 eV. Meanwhile, the stability of the Na plating/stripping behavior in symmetric cells increases from 236 to 654 h. We tried to reveal the microscopic origins of the higher ionic conductivity of MgF$_2$-doped NASICONs by comprehensive in-house characterizations. Our study discovers a novel MgF$_2$ additive and provides an efficient way to prepare higher-performance SSEs, making it possible to fabricate lower-cost SSBs in industries.
\end{abstract}

\begin{graphicalabstract}
\includegraphics[width=0.88\textwidth]{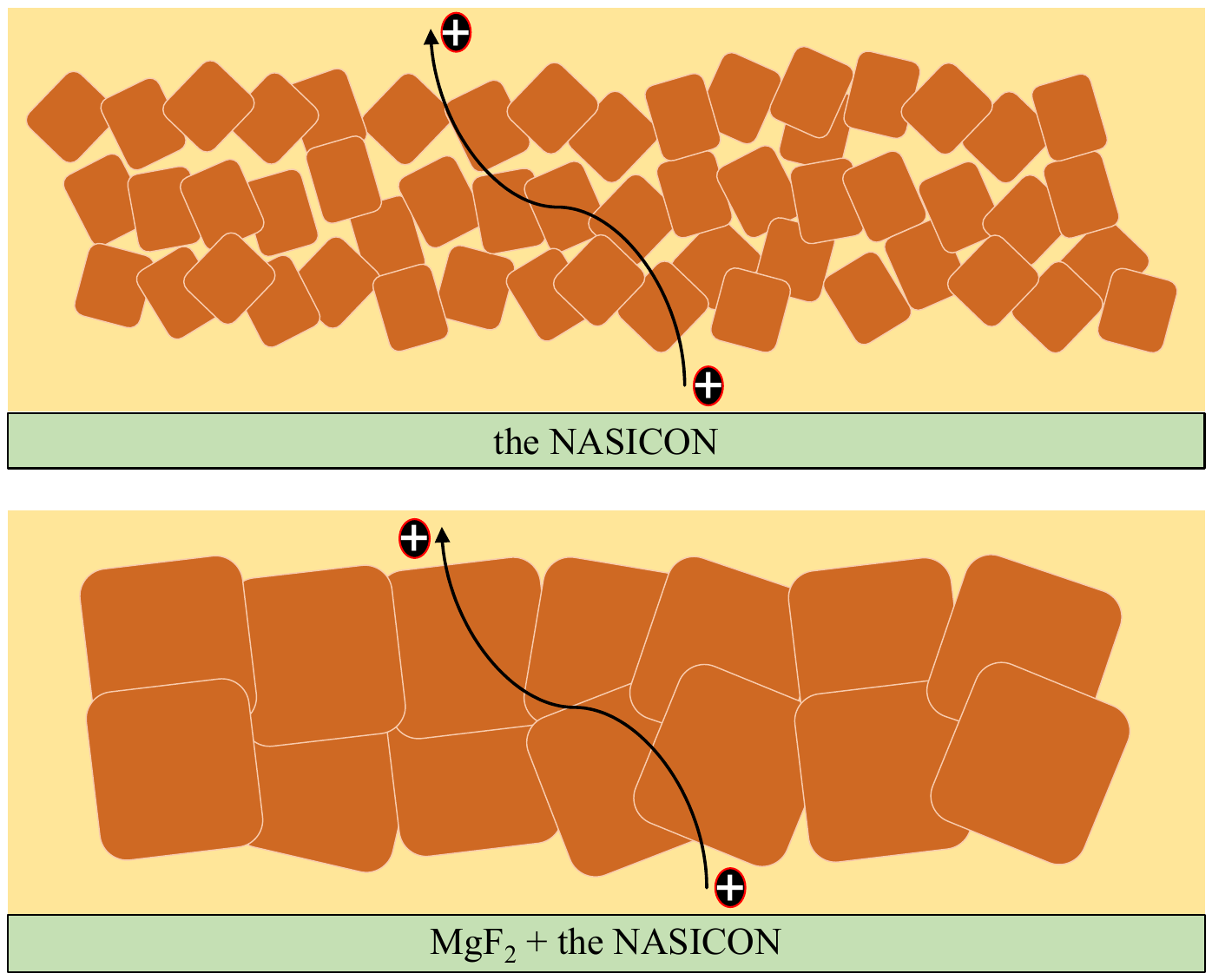}
\end{graphicalabstract}

\begin{highlights}
\item We discovered a novel additive MgF$_2$ for solid-state electrolytes.
\item MgF$_2$ additive improves the total ionic conductivity of NASICON to 2.03 mS cm$^{-2}$.
\item The Na plating/stripping stability in symmetric cells increases from 236 to 654 h.
\item Microscopic origins of the higher ionic conductivity of the MgF$_2$-doped NASICONs.
\item An efficient and universal way to prepare various higher-performance solid-state electrolytes.
\end{highlights}

\begin{keyword}
Solid-state battery \sep Solid-state electrolytes \sep NASICON \sep Ionic conductivity \sep Additive
\end{keyword}

\end{frontmatter}


\section{Introduction}
\label{Introduction}
Solid-state batteries (SSBs) hold more significantly-improved safety and much higher theoretical energy density than current liquid-based lithium-ion batteries; offering promising applications in multiple fields, especially in the large-scale energy storage system \cite{reisch2017solid, PING2019246}. Among various solid electrolytes, the family of Na$_{1+x}$Zr$_2$Si$_{x}$P$_{3-x}$O$_{12}$ ($0 \leq x \leq 3$) compounds is a typical representative of sodium-ion electrolytes. Such ionic conductor is the so-called sodium superionic conductor (NASICON) that was discovered originally by Goodenough \emph{et al}. in 1976 \cite{goodenough1976fast}, and possesses wide electrochemical window and good air stability \cite{Yang2021}. During the past, nearly, 48 years of intensive studies, new NASICON-type solid electrolytes such as Li$_{1+x}$Al$_x$Ti$_{2-x}$(PO$_4$)$_3$ (LATP) and Li$_{1+x}$Al$_x$Ge$_{2-x}$(PO$_4$)$_3$ (LAGP) have been developed for lithium-ion batteries \cite{DeWees2019}. The low conductivity of the solid electrolyte and the high interracial resistance between the electrolyte and electrodes have been the two long-standing obstacles for practical SSBs applications \cite{reisch2017solid, PING2019246, goodenough1976fast, Yang2021, DeWees2019}.

Upon cooling, NASICON undergoes a structural phase transition from a high-temperature rhombohedral $R$-3$c$ to low-temperature monoclinic $C2/c$ phase at $T_{\textrm{S}} \sim$ 431 K \cite{VONALPEN19791317, Jolley20151}. The structural transition can be tuned by chemical engineering. The transition temperature ($T_{\textrm{S}}$) was reduced by substituting aliovalent metals (Al, Y, Fe, Co, Ni, and Zn) for Zr in NASICONs. For example, $T_{\textrm{S}} =$ 412.65 K for the Y-doped NASICON \cite{Jolley20151}. When $x = 2.4$, i.e. increasing the ratio of silica and lowering the phosphorus composition, Na$_{3.4}$Zr$_2$Si$_{2.4}$P$_{0.6}$O$_{12}$ presents a room-temperature rhombohedral structure and a high bulk conductivity of 1.5 $\times$ 10$^{-2}$ S/cm that is comparable to common liquid electrolytes of sodium-ion batteries \cite{MaQ2019}. The addition of NaF into NASICON-systems results in a gradual structure transformation from a monoclinic phase of NASICON grains into a rhombohedral phase, and achieves an optimum conductivity of 3.6 $\times$ 10$^{-3}$ S{/}cm at $\sim$ 25$^{\circ}$C for the Na$_{3.2}$Zr$_2$Si$_{2.2}$P$_{0.8}$O$_{12}$-0.5NaF sample \cite{Shao2019}. To improve the ionic conductivity, divalent Mg was widely used as a doping element into NASICONs \cite{yang2018nasicon}. The general consensus reached is that Mg$^{2+}$ ions partially occupy the crystallographic position of Zr$^{4+}$, as other dopants such as Ca$^{2+}$ and Zn$^{2+}$ do \cite{Lu2019, Yang2020}. The doping of hetero-valent elements into NASICONs increases the concentration of mobile sodium ions, and thus enhances the correlated migration, while the addition of F, Mg, and La during the synthesis process promotes a formation of a second phase and tailors the stoichiometry between grain boundaries \cite{Wang2021, fu2019reducing, chen2018dielectric}. Albeit with these achievements, there are still many issues existing in the study of oxide electrolytes such as the high interfacial resistance, ceramic fragility, and dendrite growth \cite{reisch2017solid, PING2019246, goodenough1976fast, Yang2021, DeWees2019}. Presently, most of the researches focus on optimizing crystal models, improving electrochemical properties, optimizing interfacial impedance, and investigating mechanisms of ionic transportation and dendrite growth \cite{Xiao2020, Chen2020, KIM2015299, dudney2015handbook}. Additives =-  =into raw precursors were used to modify the performance of NASICONs by achieving an uniform surface and a high solidification density of electrolytes \cite{Randau2021}. These additives such as Na$_2$SiO$_3$, Na$_3$BO$_3$, Bi$_2$O$_3$, Na$_2$B$_4$O$_7$, and antimony/tin oxides can function as fluxes in a liquid phase sintering technique, lowering the sintering temperature and improving the ionic conductivity \cite{oh2019composite, miao2021influence, zhao2021homogeneous}.

Here, we report on an additive of MgF$_2$ and investigate its effect on a NASICON solid-state electrolyte. Based on the melting point of MgF$_2$, i.e. $1255 \pm 3$ $^{\circ}$C \cite{Duncanson1958}, we sintered the electrolyte at 1200--1300 $^{\circ}$C. We found that mere 1 wt$\%$ of MgF$_2$ addition can improve the ionic conductivity from 1.12 to 2.03 mS{/}cm, increased up to $\sim$ 181.3\%. Meanwhile, we have optimized the synthesis process with significantly shortened sintering time. Our research reveals that MgF$_2$ is an ideal additive for industrial production of higher-quality electrolytes with lower cost.

\begin{figure*} [!t]
\centering \includegraphics[width=0.82\textwidth]{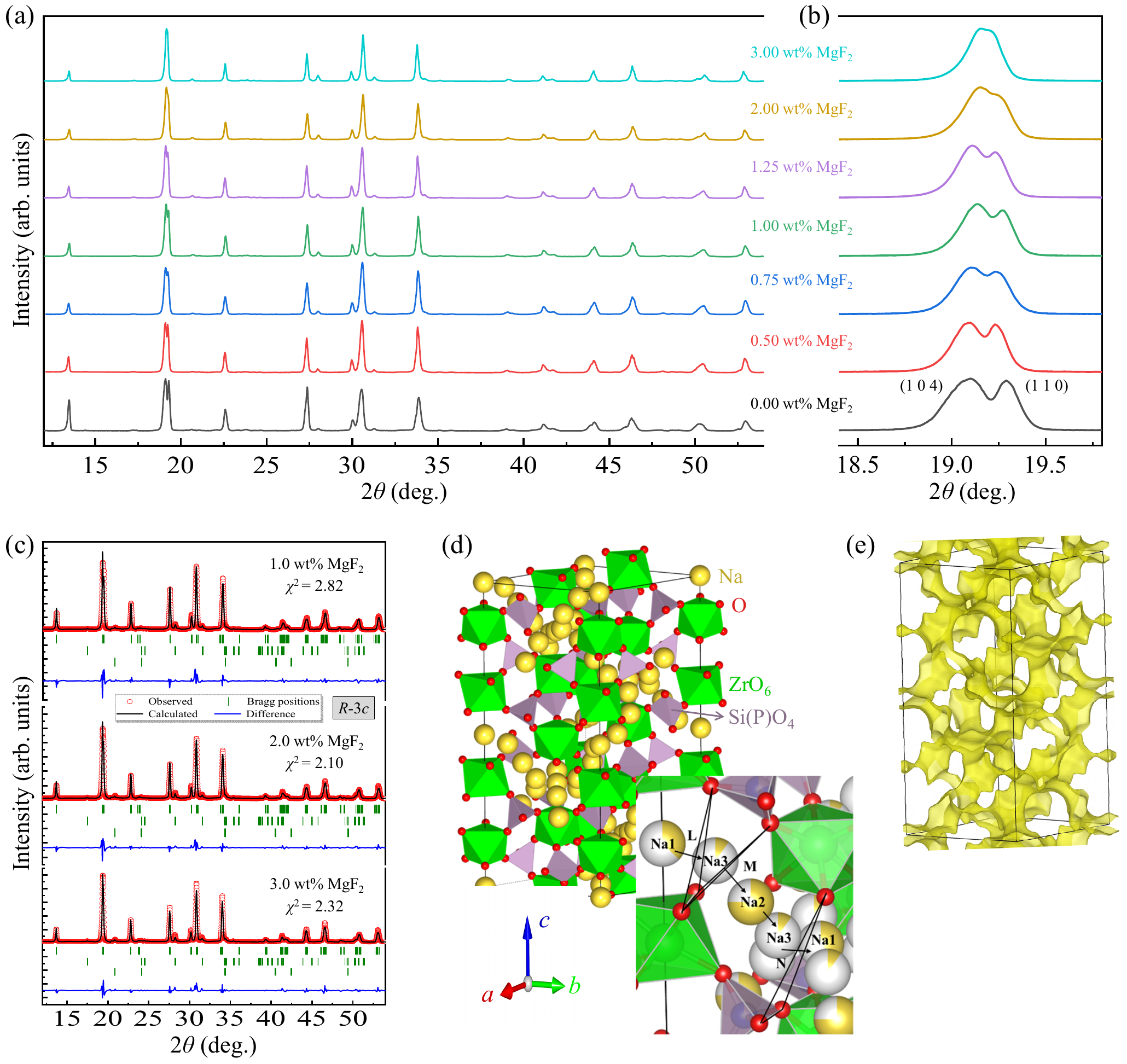}
\caption{
(a) Room-temperature X-ray powder diffraction patterns of the Na$_{3.4}$Zr$_2$Si$_{2.4}$P$_{0.6}$O$_{12}$-$z\cdot$MgF$_2$ ($z =$ 0.00, 0.50, 0.75, 1.00, 1.25, 2.00, and 3.00 wt\%) NASICONs in the 2$\theta$ range of 12--54$^{\circ}$.
(b) Corresponding to (a), 2$\theta$ is in the range of 18.4--19.8$^{\circ}$.
(c) Observed (circles) and calculated (solid lines) X-ray powder diffraction patterns of the Na$_{3.4}$Zr$_2$Si$_{2.4}$P$_{0.6}$O$_{12}$-$z\cdot$MgF$_2$ ($z =$ 1.0, 2.0, and 3.0 wt\%) NASICONs, collected at room temperature. The vertical bars mark the positions of Bragg reflections from three phases: the MgF$_2$-doped NASICONs (1$^{\textrm{st}}$ row, space group: rhombohedral $R$-3$c$), ZrO$_2$ (2$^{\textrm{nd}}$ row, space group: monoclinic $P$12$_1${/}$c$1), and Na$_3$PO$_4$ (3$^{\textrm{rd}}$ row, space group: cubic $Fm$-3$m$). The lower curves represent the difference between observed and calculated patterns. The value of ${\chi}^2$ indicates the goodness of fit.
(d) Crystal structure of the rhombohedral Na$_{3.4}$Zr$_2$Si$_{2.4}$P$_{0.6}$O$_{12}$ NASICON with an $R$-3$c$ space group. The ZrO$_6$ octahedrons (green) and the Si(P)O$_4$ tetrahedrons (purple) are displayed. The Na (Na1, Na2, and Na3) and O positions are labeled.
(e) The isosurface wave was calculated by the bond valence energy landscape, which could be taken as the ionic migration channel.
}
\label{structure}
\end{figure*}

\begin{table}[!t]
\small
\centering
\caption{Lattice constants ($a$ and $c$) and unit-cell volumes of NASICONs with different MgF$_2$-doping levels. These were extracted from Rietveld refinements of X-ray powder diffraction data collected at room temperature.}
\label{LConstants}
\setlength{\tabcolsep}{8.8mm}{}
\renewcommand{\arraystretch}{1.1}
\begin{tabular}{cccc}
\hline
\hline
MgF$_2$-doped NASICONs            & $a$ (\AA)       & $c$ (\AA)       & Unit-cell volume ({\AA}$^3$)        \\ [1pt]
\hline
0.0 wt$\%$ MgF$_2$                & 9.1010(2)       & 22.6709(9)      & 1626.22(1)                          \\
1.0 wt$\%$ MgF$_2$                & 9.1084(1)       & 22.5800(6)      & 1622.31(1)                          \\
2.0 wt$\%$ MgF$_2$                & 9.1125(1)       & 22.5459(5)      & 1621.54(1)                          \\
3.0 wt$\%$ MgF$_2$                & 9.1261(1)       & 22.5116(6)      & 1623.90(1)                          \\
\hline
\hline
\end{tabular}
\end{table}

\begin{figure*} [!t]
\centering \includegraphics[width=0.82\textwidth]{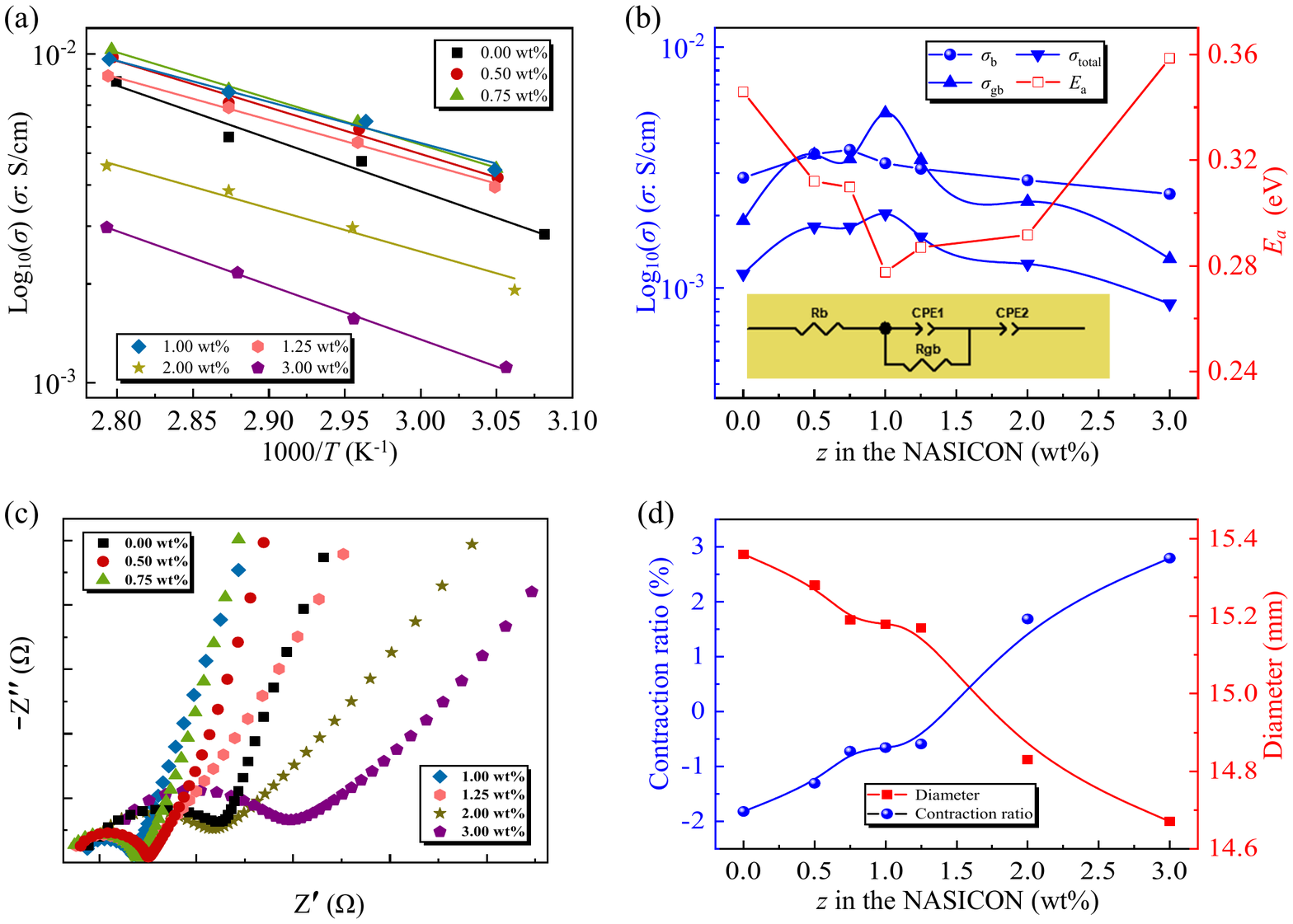}
\caption{
(a) Arrhenius plot of the ionic conductivity ${\sigma}$ for the Na$_{3.4}$Zr$_2$Si$_{2.4}$P$_{0.6}$O$_{12}$-$z\cdot$MgF$_2$ ($z =$ 0.00, 0.50, 0.75, 1.00, 1.25, 2.00, and 3.00 wt\%) NASICONs, i.e., ${\sigma}$ vs. 1000$T^{-1}$. The vertical axis is logarithmic.
(b) Calculated bulk (${\sigma}_\textrm{b}$), grain boundary (${\sigma}_\textrm{gb}$) and total ionic (${\sigma}_\textrm{total}$) conductivities (left), and estimated activation energy of conduction ($E_\textrm{a}$) (right) for the MgF$_2$-doped NASICONs. The left vertical axis is logarithmic. Inset schematically shows the A.C. equivalent circuit for the fitting, where Rb and Rgb represent resistances of the bulk and the grain boundary, respectively. The constant phase elements (CPEs) simulate the leaking capacitor model in the grain boundary (CPE1) and blocking electrodes (CPE2).
(c) Nyquist plot of the impedance of MgF$_2$-doped NASICONs measured at room temperature, that is, negative of the imaginary part (-$Z^{\prime\prime}$) vs. the real part ($Z^{\prime}$).
(d) Variation of the contraction ratio (left) and the diameter (right) of the sintered pellets of the MgF$_2$-doped NASICONs at different doping levels.
}
\label{conduct}
\end{figure*}

\begin{figure*} [!t]
\centering \includegraphics[width=0.82\textwidth]{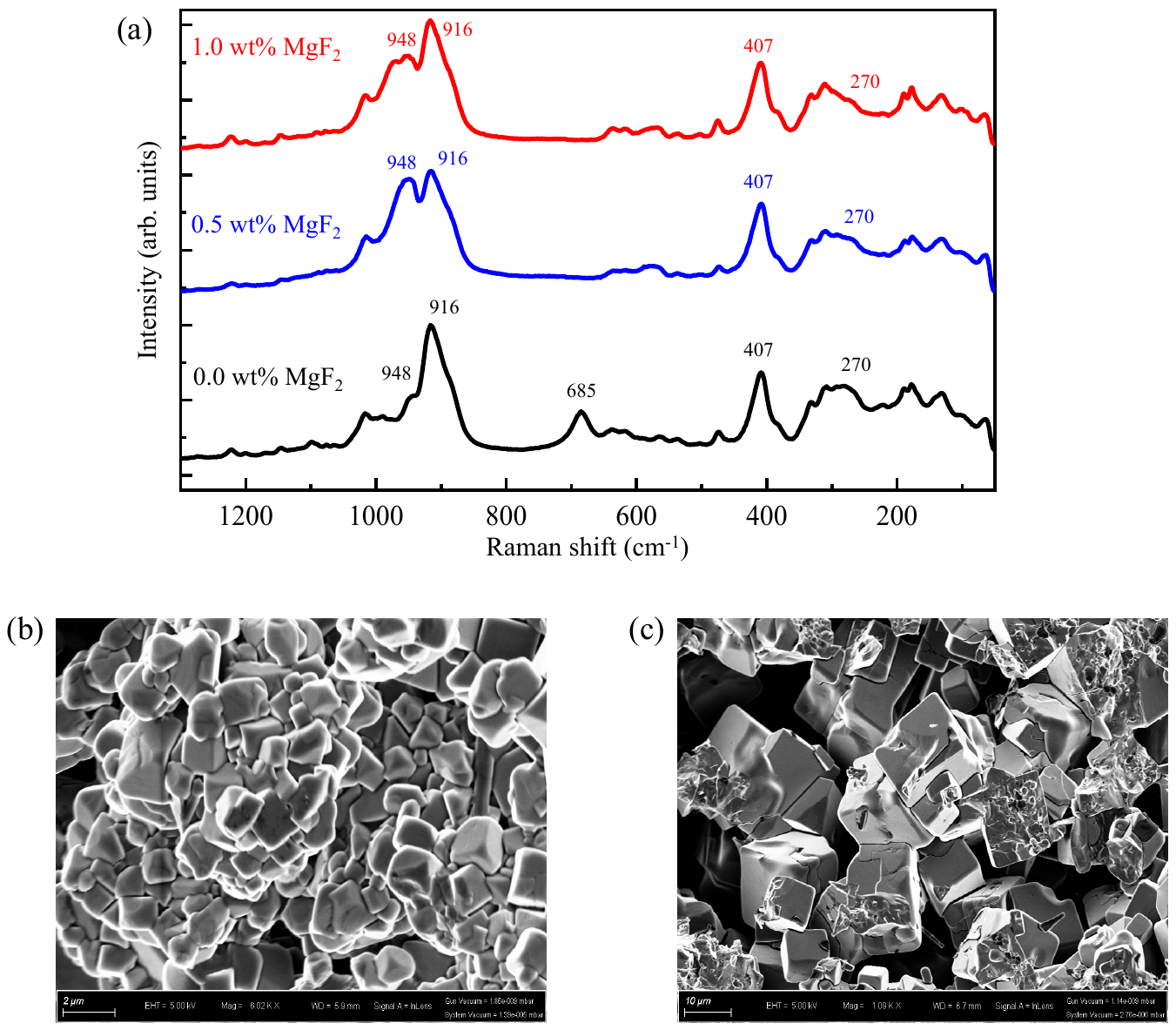}
\caption{
(a) Raman spectra of the Na$_{3.4}$Zr$_2$Si$_{2.4}$P$_{0.6}$O$_{12}$-$z\cdot$MgF$_2$ ($z =$ 0.0, 0.5, and 1.0 wt\%) NASICONs.
(b) A scanning electronic microscopy image of the Na$_{3.4}$Zr$_2$Si$_{2.4}$P$_{0.6}$O$_{12}$ NASICON particle. Scale bar represents 2 $\mu$m.
(c) A scanning electronic microscopy image of the Na$_{3.4}$Zr$_2$Si$_{2.4}$P$_{0.6}$O$_{12}$-$z\cdot$MgF$_2$ ($z =$ 1.0 wt\%) NASICON particle. Scale bar represents 10 $\mu$m.
}
\label{RSX}
\end{figure*}

\begin{figure*} [!t]
\centering \includegraphics[width=0.82\textwidth]{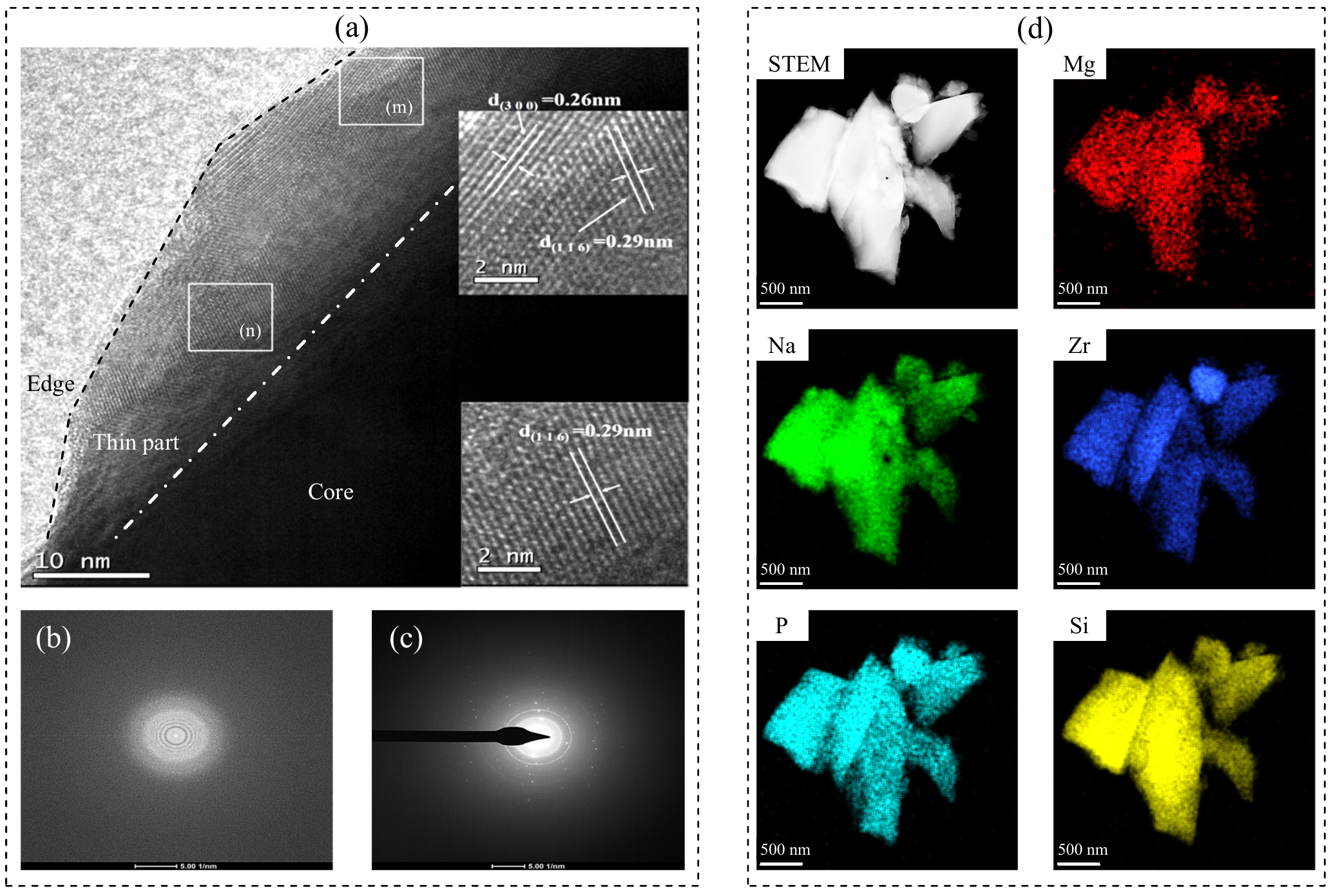}
\caption{
(a) A high-resolution transmission electron microscopy (HRTEM) image of NASICON. The marked (m) regime is indexed as Miller indices of (3 0 0) and (1 1 6), while the (n) regime corresponds to only (1 1 6) index.
(b) An electron diffraction pattern of the NASICON particle from the edge regime as marked above the dashed line in (A).
(c) An electron diffraction pattern of the NASICON particle from the thin-part regime bordering the dashed and dash-dotted lines as indicated in (A).
(d) A scanning transmission electron microscopy image of the NASICON and EDS mapping of Mg, Na, Zr, P, and Si elements.
}
\label{TEM}
\end{figure*}

\begin{figure*} [!t]
\centering \includegraphics[width=0.82\textwidth]{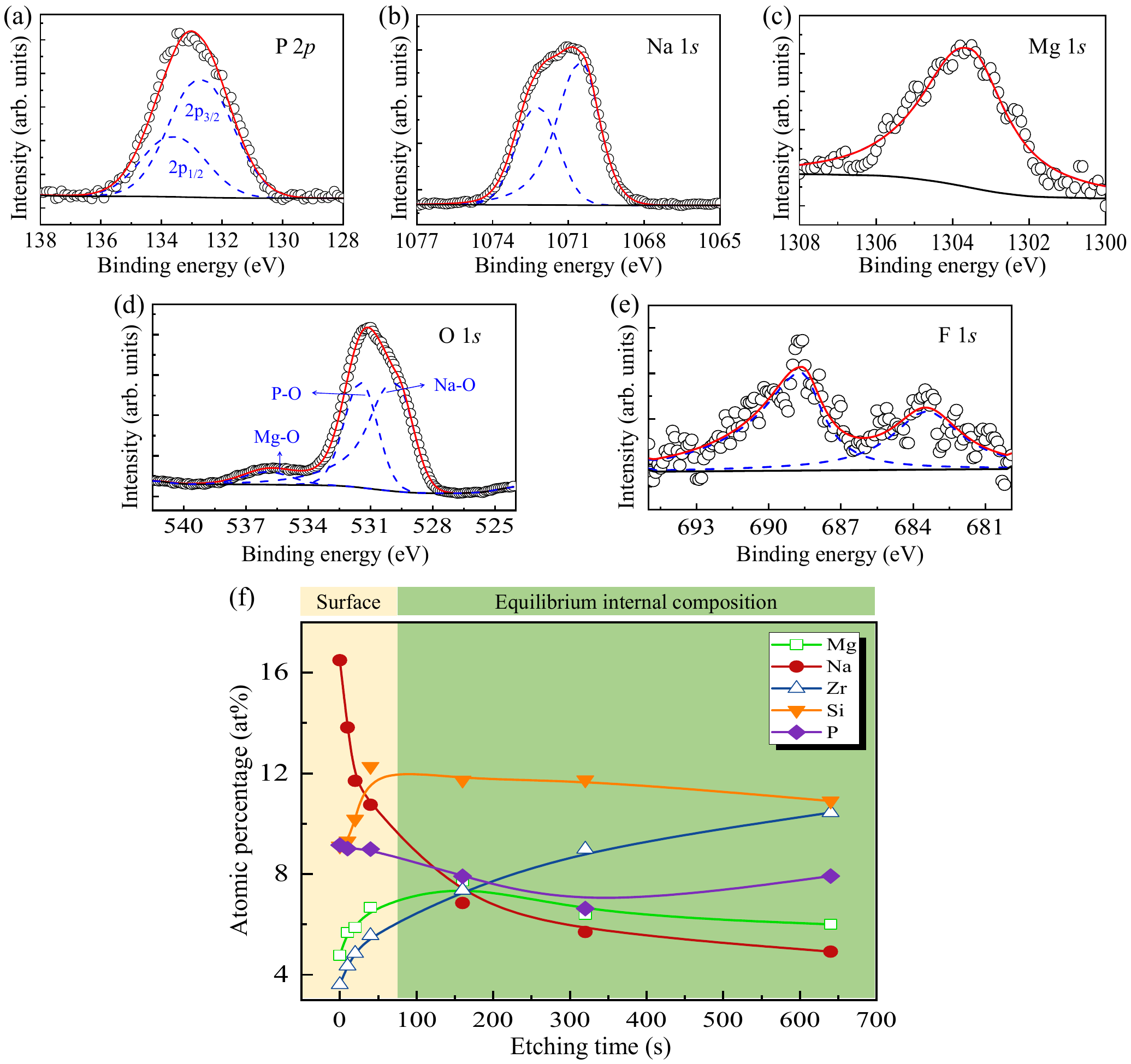}
\caption{
(a-e) X-ray photoelectron spectroscopy spectra at the (a) P 2$p$-edge, (b) Na 1$s$-edge, (c) Mg 1$s$-edge, (d) O 1$s$-edge, and (e) F 1$s$-edge for the surface of the Na$_{3.4}$Zr$_2$Si$_{2.4}$P$_{0.6}$O$_{12}$-$z\cdot$MgF$_2$ ($z =$ 1.0 wt\%) NASICON.
(f) Correlation between etching time (depth) and element compositions of the Na$_{3.4}$Zr$_2$Si$_{2.4}$P$_{0.6}$O$_{12}$-$z\cdot$MgF$_2$ ($z =$ 1.0 wt\%) NASICON. The etching speed refers to Ta$_2$O$_5$, i.e. 0.19 nm/s.
}
\label{XPS}
\end{figure*}

\begin{figure*} [!t]
\centering \includegraphics[width=0.82\textwidth]{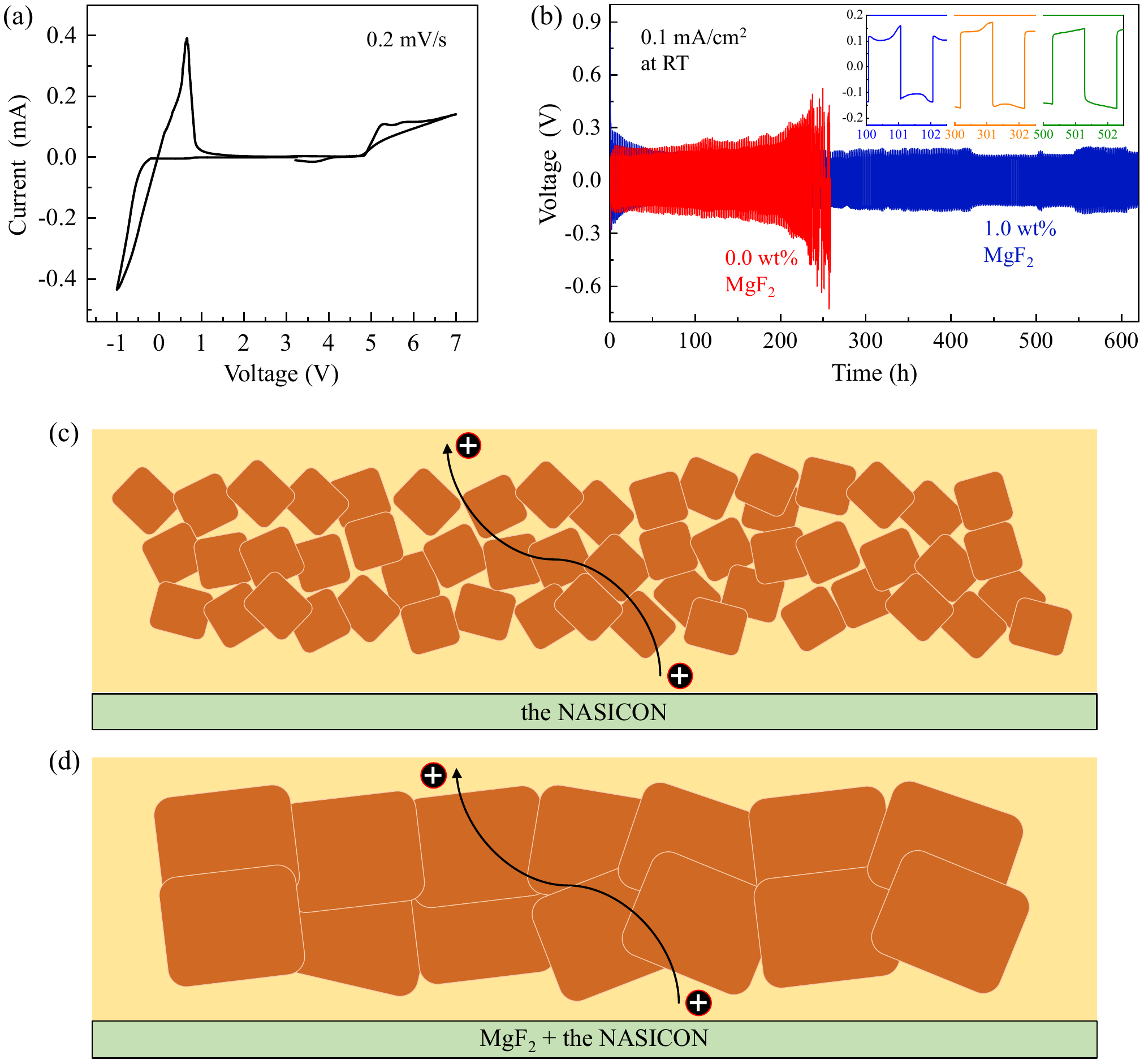}
\caption{
(a) A current-voltage curve of the Na{/}Na$_{3.4}$Zr$_2$Si$_{2.4}$P$_{0.6}$O$_{12}$-1 wt{\%}$\cdot$MgF$_2${/}stainless steel cell at a voltage change rate of 0.2 mV{/}s.
(b) Galvanic cycles at room temperature show the differences in the charge/discharge cycles between the symmetric cells of Na{/}Na$_{3.4}$Zr$_2$Si$_{2.4}$P$_{0.6}$O$_{12}${/}Na (red) and Na{/}Na$_{3.4}$Zr$_2$Si$_{2.4}$P$_{0.6}$O$_{12}$-1 wt{\%}$\cdot$MgF$_2${/}Na (blue). Inset shows the enlarged regimes within 100--102.5 h (left), 300--302.5 h (middle), and 500--502.5 h (right).
(c-d) Diagrams show schematically the difference between the (c) Na$_{3.4}$Zr$_2$Si$_{2.4}$P$_{0.6}$O$_{12}$ and (d) MgF$_2$-doped Na$_{3.4}$Zr$_2$Si$_{2.4}$P$_{0.6}$O$_{12}$ NASICONs.
}
\label{cell}
\end{figure*}

\begin{figure*} [!t]
\centering \includegraphics[width=0.64\textwidth]{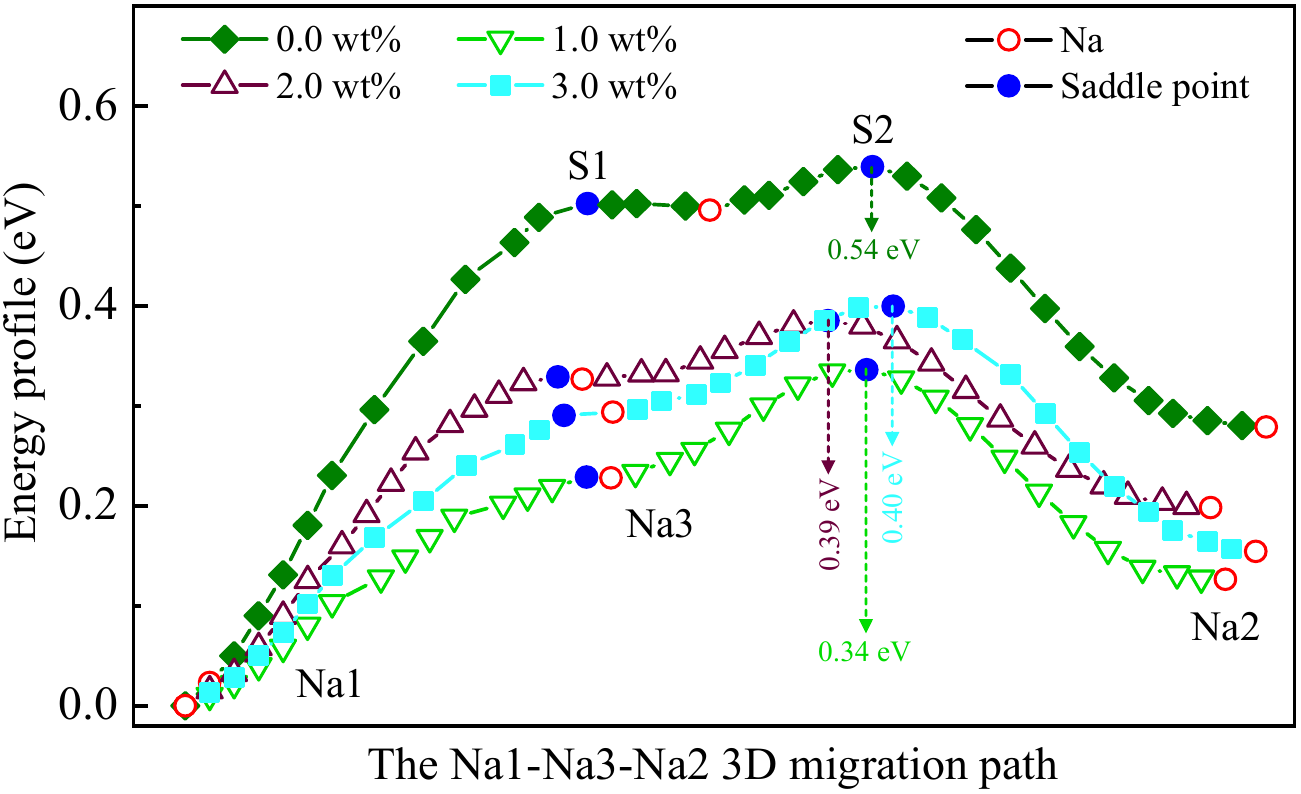}
\caption{
Based on the extracted structural information, we calculated Na$^+$ migration barriers in the NASICONs by using the software of softBV. The maximum values were marked by the vertical dashed arrows. The saddle points (solid circles) and the Na$^+$ sites (void circles) represent the possible equilibrium states and interstitial sites, respectively.
}
\label{Softbv}
\end{figure*}

\section{Materials and methods}
\label{}
\subsection*{2.1 Materials preparation}

NASICON-type Na$_{1+x}$Zr$_2$Si$_{x}$P$_{3-x}$O$_{12}$-$z\cdot$MgF$_2$ ($z =$ 0.00, 0.50, 0.75, 1.00, 1.25, 2.00, and 3.00 wt\%) samples were synthesized by a solid-state reaction method \cite{Li2008-1}. In this study, the nominal composition of the NASICON sample without additive MgF$_2$ is Na$_{3.4}$Zr$_2$Si$_{2.4}$P$_{0.6}$O$_{12}$. We first weighted the raw materials of Na$_3$PO$_4$ (Alfa Aesar, 99.99\%), Na$_2$CO$_3$ (Alfa Aesar, 99.99\%), SiO$_2$ (Alfa Aesar, 99.99\%), and ZrO$_2$ (Alfa Aesar, 99.99\%), with additional Na$_3$PO$_4$ and Na$_2$CO$_3$ compounds for 15{\%} mole each to compensate for the volatilization of Na and P elements during the syntheses. The initial mixture of all raw materials was milled and mixed with a 50 mm-diameter ball by a Vibratory Micro Mill (FRITSCH PULVERISETTE 0) for 1 h and then calcined at 1100 $^{\circ}$C for 12 h, with both heating-up and cooling-down ramps of 200 $^{\circ}$C/h under air ambient. The resultant bulk sample was pulverized and divided into different parts. We named them as parent powder samples. After mixing homogeneously some of the parent samples with different mass fractions of MgF$_2$, we pressurized the mixtures into pellets with an axial pressure of $\sim$ 800 Mpa. The pellets were buried into the corresponding powdered samples and placed into an aluminum oxide crucible for sintering \cite{narayanan2019sintering, wen2022effect}, by which the volatilization of light elements was evidently reduced during the sintering process at 1250 $^{\circ}$C. After sintering, we assure that it is easy to separate the pellet from the crucible and the powdered sample. The sintered pellets were stored in a glove box under Ar atmosphere to avoid possible reactions with CO$_2$ and H$_2$O in air.

\subsection*{2.2 X-ray powder diffraction}

The sintered pellets of the Na$_{1+x}$Zr$_2$Si$_{x}$P$_{3-x}$O$_{12}$-$z\cdot$MgF$_2$ ($z =$ 0.00, 0.50, 0.75, 1.00, 1.25, 2.00, and 3.00 wt\%) NASICONs were gently pulverized into the powder samples and characterized from 2$\theta =$ 12 to 54$^{\circ}$ with a step size of 0.01$^{\circ}$ on an in-house X-ray powder diffractometer (Rigaku, SmartLab 9 kW), employing both the copper $K_{\alpha1}$ (1.54056 {\AA}) and $K_{\alpha2}$ (1.54439 {\AA}) with a ratio of 2{:}1 as the radiation, in a Bragg-Brentano geometry at a voltage of 45 kV and a current of 200 mA under ambient conditions.

\subsection*{2.3 Crystallography}

The software FULLPROF SUITE was utilized for the crystallography refinements with collected X-ray powder diffraction data at room temperature \cite{fullprof}. We refined the background with a linear interpolation between automatically-selected points. The pseudo-voigt function was used for modeling a Bragg peak shape. We refined the scale factor, zero shift, peak shape parameters, asymmetry, lattice parameters, atomic positions, and isotropic thermal parameters.

\subsection*{2.4 Electrochemical impedance spectroscopy}

Blocking electrodes (Ag or Al) were plated on both sides of the NASICONs using a vacuum vapor deposition technique. We used conductive metal-clips to hold the prepared pellets and measured them with a voltage of 5 mV in the frequency range from 1 Hz to 1 MHz in air. An alternative to making blocking electrodes is using a conductive silver paste or sputtering gold. For both the methods, there exist two critical points: one is the close contact between blocking electrodes and pellets; the other is that the preparation process should not affect the result of measurement by electrochemical impedance spectroscopy. There are two types of conductive silver pastes: sintered silver paste and quick-drying silver paste. The organic adhesive in the quick-drying silver paste can infiltrate into the pellet, which would affect the NASICON's performance around grain boundaries. The binder removal can also affect the final performance of the electrolyte.

\subsection*{2.5 Raman spectrum}

The surfaces of the Na$_{3.4}$Zr$_2$Si$_{2.4}$P$_{0.6}$O$_{12}$-$z\cdot$MgF$_2$ ($z =$ 0.0, 0.5, and 1.0 wt\%) NASICONs were characterized by Raman scattering on a confocal laser Raman spectrometer (LabRAM HR Evolution, Horiba, Japan), with a 532 nm solid-state laser as an excitation source. A 50 $\times$ magnification of optical lens was used for Raman measurements. The acquisition time and the accumulation time are both 5 s for the Raman measurements.

\subsection*{2.6 Scanning electron microscopy}

Scanning electron microscopy images of the Na$_{3.4}$Zr$_2$Si$_{2.4}$P$_{0.6}$O$_{12}$ and Na$_{3.4}$Zr$_2$Si$_{2.4}$P$_{0.6}$O$_{12}$-$z\cdot$MgF$_2$ ($z =$ 1.0 wt\%) NASICONs were measured with ZEISS Sigma. Both the samples were sputter-coated with gold to make them electrically conductive. This study was carried out in high vacuum under an accelerating voltage of 5 kV.

\subsection*{2.7 X-ray photoelectron spectroscopy}

We characterized the elemental contents and chemical properties both on the surface and inside of the pellet of the Na$_{3.4}$Zr$_2$Si$_{2.4}$P$_{0.6}$O$_{12}$-$z\cdot$MgF$_2$ ($z =$ 1.0 wt\%) NASICON with X-ray photoelectron spectroscopy (XPS) (Thermo Fisher Scientific, ESCALAB 250Xi), with incident energy of 50.0 eV for a full-scan spectrum and 20.0 eV for a narrow-scan spectrum, with an energy step size of 0.1 eV.

\subsection*{2.8 Scanning transmission electron microscopy}

The powdered samples of the Na$_{3.4}$Zr$_2$Si$_{2.4}$P$_{0.6}$O$_{12}$ and Na$_{3.4}$Zr$_2$Si$_{2.4}$P$_{0.6}$O$_{12}$-$z\cdot$MgF$_2$ ($z =$ 1.0 wt\%) NASICONs were dispersed in ethanol with ultrasonic treatment for 5 min. The resulting individual particles were characterized by scanning transmission electron microscope (FEI, Talos F200X) at 200 kV. The elemental compositions as well as their distributions were analyzed by high-resolution imaging with energy dispersive X-ray spectroscopy.

\subsection*{2.9 Cell assembly}

The symmetric cells were assembled in a Swagelok cell. The NASICON-type pellets were infiltrated into a liquid electrolyte, after which the excess electrolyte was adsorbed by dust-free papers. The liquid electrolyte fills the pores up inside ceramics, which can reduce the formation of passivation layers between the ceramic and sodium metal \cite{gao2018stabilizing}. The sodium metal was first cut into thin slices and put into a plastic bag in the Ar-filled glove box. The sodium-slices were then pressed into much thinner foils using a glass rod, and the round sodium foil was removed by a hole punch. We chose good sodium foils and attached them on both sides of the NASICON pellets. This assembly of the NASICON pellet and sodium-foils was sealed in the Swagelok cell. The whole battery assemblies were carried out in the glove box with Ar atmosphere.

\section{Results and discussion}

As shown in Fig.~\ref{structure}a, we studied the crystallographic structure and phase purity of the Na$_{3.4}$Zr$_2$Si$_{2.4}$P$_{0.6}$O$_{12}$-$z\cdot$MgF$_2$ ($z =$ 0.00, 0.50, 0.75, 1.00, 1.25, 2.00, and 3.00 wt\%) NASICONs by X-ray powder diffraction. Fig.~\ref{structure}b shows the evolution of Bragg (104) and (110) peaks. For the Na$_{3.4}$Zr$_2$Si$_{2.4}$P$_{0.6}$O$_{12}$ sample, peak positions of the Bragg (104) and (110) reflections locate at $\sim$ 19.08$^{\circ}$ and 19.31$^{\circ}$, respectively, with an intensity ratio of \emph{I}$_{\textrm{(104)}}${/}\emph{I}$_{\textrm{(110)}}$ $\approx$ 2.32. As the additive level $z$ increases, the two Bragg peaks get closer and closer and eventually merge into a quasi-single peak at $z =$ 3.00 wt\%. The evolutions of peak position and intensity with \emph{z} indicate changes of crystallography in lattice constants and atomic positions. To quantitatively evaluate these parameters, we refined the collected X-ray powder diffraction patterns of the Na$_{3.4}$Zr$_2$Si$_{2.4}$P$_{0.6}$O$_{12}$-$z\cdot$MgF$_2$ ($z =$ 0.00, 1.00, 2.00, and 3.00 wt\%) NASICONs (Fig.~\ref{structure}c), and the extracted lattice parameters are listed in Table~\ref{LConstants}. The resulting atomic positions are listed in Table S1. By increasing the $z$ value, the lattice constants \emph{a} and \emph{c} almost increase and decrease linearly, respectively, which jointly results in a valley bottom for the unit-cell volume around $z =$ 1.75 wt\% (Fig.~S1).

The Bragg peaks located at 2$\theta$ $\sim$ 28.25$^{\circ}$ and 31.54$^{\circ}$ originate from an ZrO$_2$ impurity (Fig.~\ref{structure}c). This was caused by the volatilization of the Na and P elements during the sintering process. We also observed impurity Bragg peaks (whose 2$\theta$ $\sim$ 17.48$^{\circ}$, 20.92$^{\circ}$, 23.92$^{\circ}$, and 24.11$^{\circ}$) corresponding to the Na$_3$PO$_4$ compound (Fig.~\ref{structure}c). Thus, we refined the X-ray powder diffraction patterns with three phases and found that the ZrO$_2$ and Na$_3$PO$_4$ phases account for $\sim$ 1.59$\%$ and 1.36$\%$ in the 1.00 wt$\%$$\cdot$MgF$_2$-doped NASICON sample, respectively. In addition, we did not find MgF$_2$ additive in all the collected patterns.

The Na$_{3.4}$Zr$_2$Si$_{2.4}$P$_{0.6}$O$_{12}$ sample crystallizes into the rhombohedral structure with an $R$-3$c$ space group at room temperature. In this study, the doping with MgF$_2$ additive up to a level of 3.00 wt\% did not induce the structure phase transition from rhombohedral to monoclinic ($C2/c$). Compared to the monoclinic phase, the rhombohedral phase has more continuous ionic channels and lower energy barrier for the Na$^{+}$ migration between crystallographic sites (Fig.~\ref{structure}d). The gradual integration of the Bragg (104) and (110) peaks (Fig.~\ref{structure}b) implies that MgF$_2$ additive strengthens the crystallinity of the rhombohedral phase, and thus reduces the activation energy \cite{deng2018crystal, monchak2016lithium, catti2000lithium}.

As shown in the right-bottom of Fig.~\ref{structure}d, the migration channel of Na$^{+}$ ions is through Na1-Na3-Na2-Na3-Na1 chains in the rhombohedral structure. Bottlenecks L, M, and N (as marked in Fig.~\ref{structure}d) control the Na1-Na3, Na3-Na2, and Na3-Na1 channels, respectively \cite{zou2021identifying, boilot1987crystal}. The areas of the bottlenecks L, M, and N in the parent Na$_{3.4}$Zr$_2$Si$_{2.4}$P$_{0.6}$O$_{12}$ sample are $\sim$ 5.109 {\AA}$^2$, 5.005 {\AA}$^2$, and 4.030 {\AA}$^2$, respectively. When 1.00 wt\% MgF$_2$ was doped into the parent sample, the areas of the bottlenecks L, M, and N expanded to $\sim$ 5.695 {\AA}$^2$, 5.247 {\AA}$^2$, and 4.442 {\AA}$^2$, respectively. When $z$ = 2.00 wt\%, the bottlenecks have been slightly retracted to $\sim$ 5.442 {\AA}$^2$, 5.131 {\AA}$^2$, and 4.433 {\AA}$^2$, respectively, but still larger than those in the parent sample. When $z$ = 3.00 wt\%, the bottlenecks change into $\sim$ 5.545 {\AA}$^2$, 5.196 {\AA}$^2$, and 4.271 {\AA}$^2$, respectively.

The ionic conductivities ($\sigma$) (Figs.~\ref{conduct}a and b) at different temperatures have been investigated by electrochemical impedance spectroscopy (EIS) (Fig.~\ref{conduct}c). The electrochemical impedance spectrum resulting from the A.C. equivalent circuit (inset of Fig.~\ref{conduct}b) was fitted by the software of ZView2 \cite{jin2013li7la3zr2o12, arbi2015high}, and the conductivity was calculated by
\begin{equation}
\sigma = \frac{1}{R}\frac{L}{S},
\label{conductivitycalculate}
\end{equation}
where \emph{R} represents the total resistance, i.e. the sum of both bulk and grain-boundary resistances, \emph{L} is the thickness of the sample pellet, and \emph{S} is the area of the blocking electrode. Limited by the frequency range of the impedance spectroscopy measuring system (below 1 MHz), the half of the semi-circle of a grain boundary in a Nyquist plot was displayed only (Fig.~\ref{conduct}c). The total resistance of the electrolyte inferred from the intersection of the semi-circle and horizontal axis results from the grains as well as their boundaries. The bare sample without MgF$_2$ additive has $\sim$ 89.58 $\Omega$ of the overall pellet impedance (black square). The lowest impedance of the overall pellet is $\sim$ 54.24 $\Omega$ for the $z$ = 1 sample (blue rhombus). The impedance directly obtained from Nyquist plots was affected by the thickness and electrode area of the pellets. The fitted impedance was calculated by the Eq.~(\ref{conductivitycalculate}).

As shown in Fig.~\ref{conduct}a, the logarithm of the conductivity changes linearly as a function of 1/$T$ over the temperature range of 25-85 $^{\circ}$C. The Arrhenius equation can be expressed as
\begin{equation}
\sigma={\sigma_0} {T^{-1}} e^{-E_a/k_BT},
\label{Arrhenius}
\end{equation}
where $k_B$ = 1.38062 $\times$ 10$^{-23}$ J/K is the Boltzmann constant, $E_a$ is the activation energy, and $\sigma_0$ is a pre-factor, therefore, the activation energy could be deduced \cite{famprikis2019fundamentals, rao2021review, zhang2017self}.

Fig.~\ref{conduct}b shows the conductivity data of $\sigma_{\textrm{total}}$, $\sigma_{\textrm{b}}$, and $\sigma_{\textrm{gb}}$ of the MgF$_2$ added NASICON samples. These were measured at 25 $^{\circ}$C and derived by the foregoing discussed method. With increasing amount of MgF$_2$ additive, the activation energy gradually decreases until $z$ = 1.00. The bare sample ($z$ = 0.00) has an activation energy of 0.311 eV, and the NASICON sample with $z$ = 1.00 has the lowest activation energy, i.e. 0.277 eV. Whereas the $\sigma_{\textrm{total}}$ and $\sigma_{\textrm{gb}}$ conductivities exhibit broad peaks around $z$ = 1.00, opposite the activation energy dependence of the $z$ value. A comparison of ionic conductivities with the areas of the bottlenecks L, M, and N clearly shows that they display a similar trend with respect to the $z$ value. Therefore, MgF$_2$ additive doped in the parent NASICON during the sintering process can optimize the Na$^{+}$ migration channels (Fig.~\ref{structure}e) and increase the ionic conductivity.

As shown in Fig.~\ref{conduct}b, the bulk and grain-boundary conductivities for the bare sample Na$_{3.4}$Zr$_2$Si$_{2.4}$P$_{0.6}$O$_{12}$ are 2.58 and 1.99 mS/cm at 25 $^{\circ}$C, respectively. When $z$ = 1, the bulk and grain-boundary conductivities reach 3.3 mS/cm (increased by $\sim$ 0.72) and 5.28 mS/cm (increased by $\sim$ 3.29), respectively, and the total conductivity reaches 2.03 mS/cm. When $z \ge 1 $, the total conductivity decreases, and the activation energy increases slightly. These were attributed to that the excess additive of MgF$_2$ does not react with the NASICON completely during the sintering process, and the remaining unreacted impurities will lower the ionic conductivity. The bottleneck analysis show a good agreement with EIS results, which indicate that the best micro and macro performances are from the $z =$ 1 NASICON sample in our study. The biggest value of the bulk conductivity is 3.74 mS/cm for the NASICON sample at $z$ = 0.75, while the lowest bulk conductivity is 2.46 mS/cm when $z$ = 3. It is evident that the inclusion of MgF$_2$ additive does indeed results in a limited improvement of the bulk conductivity.

As shown in Fig.~\ref{conduct}d, the contraction ratio of the diameter almost increases linearly as a function of the additive level $z$. The bare sample Na$_{3.4}$Zr$_2$Si$_{2.4}$P$_{0.6}$O$_{12}$ displays a negative contraction ratio due to the volume expansion of the ceramic. When $z$ = 2.0 and 3.0, the obtained NASICON pellets show high contraction ratios of $\sim$ 1.68$\%$ and $\sim$ 2.79$\%$, respectively. Our study indicates that the hardness of the NASICON samples with excessive MgF$_2$ additive, e.g. when $z$ = 2.0 and 3.0, is increased obviously, and the corresponding NASICON samples are hard to be polished. These observations demonstrate that MgF$_2$ additive improves the sintering process of the NASICON samples. Meanwhile, the sample density releases the connection with the ionic conductivity.

To reveal the origin of conductivity enhancement achieved by using MgF$_2$ additive in the NASICONs, we have carried out Raman spectroscopy studies as shown in Fig.~\ref{RSX}a. The band shift at 685 cm$^{-1}$ was attributed to the impurity of ZrO$_2$. The several obvious band shifts at 100--300 cm$^{-1}$ originate from the bending of P-O-P. The bent vibrating and stretching modes of SiO$_4$ and PO$_4$ tetrahedra were reflected by the band shifts at 948, 916, and 407 cm$^{-1}$. The band shift around 270 cm$^{-1}$ comes from the ZrPO$_4$($/$SiO$_4$) compounds. As the $z$ value increases, the intensity of the 270 cm$^{-1}$ band-shift decreases. Correlating the MgF$_2$ doping dependence of both the bulk and grain-boundary conductivities, we attribute the conductivity enhancement to the Mg$^{2+}$ doping in the NASICON crystalline lattice, though the doping level was modest. Therefore, the doping behavior can increase the sodium concentration in the NASICON structure, which increases the atomic occupancy of Na$^+$ sites. The extra sodium sites may provide fast ion position exchange and promote the correlated migration \cite{colomban1986orientational, zhang2019correlated, deng2020phase}.

Figs.~\ref{RSX}b and c show the morphologies of the bare Na$_{3.4}$Zr$_2$Si$_{2.4}$P$_{0.6}$O$_{12}$ and Na$_{3.4}$Zr$_2$Si$_{2.4}$P$_{0.6}$O$_{12}$-1.0$\cdot$MgF$_2$ NASICONs. As the $z$ value increases from 0.0 to 1.0, the grain size of the NASICON sample increases from $\sim$ 1 to 10 $\mu$m. The fracture surface of the additive electrolyte shows features of a quasi-cleavage brittle fracture and a dimple fracture with glass shape (Fig.~S2), and most of the fractures of the bare sample are particle accumulation without close contact between particles and glassy grain boundary materials.

To further understand the microstructure of the MgF$_2$-doped NASICONs, we dispersed the Na$_{3.4}$Zr$_2$Si$_{2.4}$P$_{0.6}$O$_{12}$-1.0$\cdot$MgF$_2$ NASICON powder in anhydrous ethanol and sonicated the mixture for 5 min. The resultant separated micro-particles were used for transmission electron microscopy studies. As shown in Figs.~\ref{TEM}a-c, HRTEM images illustrate the core section (dark part), thin section (grey part), and edge section (white part) of the NASICON particles. We observed a mismatch behavior in the thin section with uniform lattice fringes of 0.29 and 0.26 nm, corresponding to Bragg (1 1 6) and (3 0 0) scattering planes, respectively. Electron diffraction patterns, as shown in Fig.~\ref{TEM}c, display characteristic features of polycrystalline materials. We observed diffraction patterns of the amorphous phases at the edge of grains and grain boundaries (Fig.~\ref{TEM}c). These disordered parts present an area where the redistribution of the mobile sodium ions occurs. Nano-sized crystals with resolved lattice fringes were also observed in the edge section (Fig.~S3) \cite{LiuYL2019}. EDS chemical mapping shows a homogeneous distribution of the Mg element in the whole particles with mere 0.3 at$\%$, and there is no detectable F element (Fig.~\ref{TEM}d). With the reported Mg-doping strategy, the Mg content should reach 0.9 at$\%$ for the highest ionic conductivity. In our work, the prepared NASICON pellets with the MgF$_2$ additive strategy reach the highest ionic conductivity with one-third amount of the equivalent Mg content. Normally, with the Mg-doping strategy, one can increase the sodium content and promote the structure transition without affecting the densification. Whereas, with the MgF$_2$ additive strategy, both effects can be generated. This indicates that the enhanced performance of MgF$_2$ is a jointed result by the synergy of several factors.

In order to learn the composition and MgF$_2$ additive distributions in the NASICON pellets, we carried out high-resolution XPS studies as shown in Figs.~\ref{XPS}a-e. The characteristic XPS peaks of P 2$p$, Na 1$s$ and Mg 1$s$ center around 133.5 eV (Fig.~\ref{XPS}a), 1071.7 eV (Fig.~\ref{XPS}b), and 1303.7 eV (Fig.~\ref{XPS}c), respectively. As to the XPS spectrum of O 1$s$, we fit the data with three peaks that were positioned at 529, 532, and 536 eV (Fig.~\ref{XPS}d). These binding energies correspond to different chemical environments of the Mg-O, P-O, and Na-O bonds, respectively. There exist two peaks around 688.6 and 683.5 eV in the F 1$s$ XPS spectrum (Fig.~\ref{XPS}e). The binding energy corresponds to that F$^-$ ions occupy O$^{2-}$ positions of the SiO$_4$ and PO$_4$ tetrahedrons \cite{he2020enhanced, zhang2020na3zr2si2po12}. These F 1$s$ peaks detected by XPS are the evidence for the existence of fluoride. Although, the signal of F 1$s$ was weak, it may indicate that the F$^-$ ions were doped into the crystalline structure.

During XPS measurements, we etched the Na$_{3.4}$Zr$_2$Si$_{2.4}$P$_{0.6}$O$_{12}$-$z\cdot$MgF$_2$ ($z =$ 1.0 wt\%) NASICON pellet with Argon beam for 0, 10, 20, 40, 160, 320, and 640 s and collected the corresponding XPS spectra after each etching in order to investigate space distributions of the NASICON compositions and MgF$_2$ additive. As shown in Fig.~\ref{XPS}f, the Mg, Na, Zr, Si, and P compositions within the surface (etching time less than 80 s) display significant changes with etching time. The Na element has the highest proportion up to 16.49 at$\%$ on the surface, and drops sharply down to 6.85 at$\%$ after 640 s etching. By contrast, the Zr element has the lowest proportion on the surface. As the etching time increases to 640 s, the Zr proportion increases up to 10.44 at$\%$. The Si composition sharply increases with the increasing etching time from 0 to 40 s and finally gets stable ($\sim$ 10 at$\%$) after 100 s etching. As to the P composition, it seems to decrease smoothly from 9.15 at$\%$ (the etching time = 0 s, surface) to $\sim$ 7 at$\%$ at the equilibrium state. The Mg composition gradually increases from 4.78 at$\%$ (the etching time = 0 s, surface) to 7.73 at$\%$ (the etching time = 160 s), followed by a smooth decrease down to 6 at$\%$ at the etching time = 640 s (Figs.~\ref{XPS}f and S4A). By contrast, the XPS intensity of the F element is too weak to be quantitatively detected (Fig.~S4B). During the sintering process, MgF$_2$ additive stays in a flowing liquid state, making it easy to diffuse through the NASICON pellet. The element compositions are more stable in an internal equilibrium state, by contrast, there exist sharp changes within the surface state.

We analyzed the electrochemical window and stability against the Na-metal electrode for the Na$_{3.4}$Zr$_2$Si$_{2.4}$P$_{0.6}$O$_{12}$-$z\cdot$MgF$_2$ ($z =$ 1.0 wt\%) NASICON by cyclic voltammetry in a Na/NASICON-1 wt$\%$$\cdot$MgF$_2$/stainless steel in a voltage range from -1 to 7 V, at a scanning rate of 0.2 mV/s at room temperature (Fig.~\ref{cell}a). There exist a couple of oxidation and reduction peaks around 0 V, which were attributed to the sodium metal plating and stripping. There is a small reduction peak at 3.60 V, corresponding to O$^{2-}$/O$_2$, leading to a leakage of O$_2$ gas. The final oxidation peak locates at 4.80 V (reference vs. Na$^+$/Na), and the decomposition products are O$_2$, SiO$_2$, Zr$_2$P$_2$O$_9$, and ZrSiO$_4$. Compared to the previously-reported wide electrochemical windows of improved NASICON electrolytes, redox reactions near low voltages are closer to real properties of NASICONs \cite{zhang2020na3zr2si2po12}. Besides that, no obvious currents due to electrolyte decomposition were observed in a scanned voltage range indicative of a stable chemical property of the Na$_{3.4}$Zr$_2$Si$_{2.4}$P$_{0.6}$O$_{12}$-1.0 wt\%$\cdot$MgF$_2$ NASICON at room temperature \cite{hartmann2013degradation}. The highly stable chemical property and the high ionic conductivity are the two main features of ceramic-oxide electrolytes. Both are the key factors for achieving SSBs with higher quality.

As shown in Fig.~\ref{cell}b, the cycling performances of two symmetric cells of Na/MgF$_2$-NASICON/Na and Na/NASICON/Na were studied at a current density of 0.10 mA/cm$^2$ at 25 $^{\circ}$C. Both symmetric cells were assembled with the same procedure. The Na/NASICON-1.0 wt$\%$$\cdot$MgF$_2$/Na has a better cycle stability at a current density of 0.1 mA/cm$^2$, indicating that the MgF$_2$-doped NASICON displays a better ability of transferring Na$^+$ ions. At the beginning of the galvanostatic cycle process, the high polarization voltage was raised to $\sim$ 0.407 V. After 20 h cycling, the low polarization voltage stabilized at 0.104 V in the Na/MgF$_2$-NASICON/Na cell. After 654 h cycling, the voltage gradually rose to 0.26 V and dropped down to 0.004 V at 688 h, indicating that a short circuit happened due to possible dendrite penetrating through the electrolyte. By contrast, the short circuit occurs at 236 h for the Na/NASICON/Na cell. According to a previous study \cite{ping2020reversible}, the ratio of ionic conductivity and its own electronic conductivity is an important factor for the dendrite growth. The larger grains with less defects of a NASICON pellet will reduce the number of sites for sodium-ion deposition into the pellet, which is in agreement with the long cycling performance of the Na/MgF$_2$-NASICON/Na cell \cite{wang2022effective}. MgF$_2$ additive appears to be in a semi-liquid state during sintering, which is like the case of so-called liquid-phase-assisted sintering. Therefore, MgF$_2$ additive in the parent NASICON can improve the driving force of sintering. Pores generated by ceramic sintering are considered as macro-defects that can lower the grain-boundary conductivity and provide more sites for dendrite growth. The semi-liquid state of MgF$_2$ additive can reduce the amount of pores during the sintering process.

Besides the densification of pellets, the obvious enhancement of the grain-boundary ionic conductivity could be also ascribed to the increase of average grain size in the MgF$_2$-doped NASICONs. As schematically illustrated in Figs.~\ref{cell}c and d, the reduced contact area between grains within a certain amount of electrolyte can efficiently lower the area of grain boundaries that are the pathway of ionic migration.

The bond valence path analyzer (BVPA) is an automated pathway analysis tool based on the software of softBV. The low computational cost and highly similar results with DFT calculation make it easy as a simplified experimental results verification tool \cite{wong2021bond, chen2019softbv}. Fig.~\ref{Softbv} shows the energy profiles of the Na$_{3.4}$Zr$_2$Si$_{2.4}$P$_{0.6}$O$_{12}$-$z\cdot$MgF$_2$ ($z =$ 0.0, 1.0, 2.0, and 3.0 wt{\%}) NASICONs, calculated by the BVPA with an energy resolution of 0.1 eV. The saddle point stays at the highest energy point, and the sodium sites represent sodium atomic positions on the possible path. Our calculation shows that the parent Na$_{3.4}$Zr$_2$Si$_{2.4}$P$_{0.6}$O$_{12}$ NASICON holds the highest energy barrier of 0.54 eV (S2). For the MgF$_2$-doped NASICONs, the energy barrier at S2 decreases significantly down to 0.40 eV ($z =$ 3.0 wt{\%}), 0.39 eV ($z =$ 2.0 wt{\%}) and 0.34 eV ($z =$ 1.0 wt{\%}). It is pointed out that the calculation only takes into account the structural information, overlooking the grain boundary and the overall difference of the pellets. We found that the trend of calculated energy barriers is consistent with that of experimentally determined activation energies. We demonstrated that MgF$_2$ additive modifies the structure of NASICONs, by which the activation energy was lowered.

\section{Conclusions}

We have studied the effect of MgF$_2$ additive on the performance of NASICONs that tend to crystallize systematically into a rhombohedral system at room temperature. Only 1 wt$\%$ MgF$_2$ additive in the Na$_{3.4}$Zr$_2$Si$_{2.4}$P$_{0.6}$O$_{12}$ NASICON results in the maximum bottleneck of the migration channel from a bond-valence energy landscape but without obvious modifications in the crystalline structure. Our SEM study shows that the grain size of the MgF$_2$-doped NASICONs increases significantly. Our EIS measurements present the variation of the grain-boundary ionic conductivity as a function of MgF$_2$ adding level. Based on the refined crystalline structural information, we calculated the energy profile by BVPA, which was in consistent with the observed trend of activation energies. Our cyclic voltammetry study presents the voltage window of the MgF$_2$-doped solid electrolytes, indicating that the doped NASICONs display similar electrochemical stability. The Galvanostatic cycling measurements of the symmetric Na/NASICONs-1.0 wt$\% \cdot$ MgF$_2$/Na cells demonstrate a long-cycle performance of 688 h. The stable plating/stripping behavior of the Na metal in the MgF$_2$-doped NASICONs implies a beneficial effect on the performance enhancement of the solid-state electrolytes. Our discovery of the novel and effective MgF$_2$ additive in NASICONs could be a universal additive for improving the performance of various ceramic electrolytes.

\section*{AUTHOR CONTRIBUTIONS}

P.F.Z., K.T.S., S.P.J., Z.R.Z., Y.F., J.C.X., S.W., and Y.H.Z. synthesized materials and carried out the measurements.
All authors discussed and analyzed the results.
P.F.Z. and H.-F.L. performed figure development and wrote the main manuscript text.
All authors reviewed the paper.
P.F.Z., H.K.N., and H.-F.L. conceived and directed the project.

\section*{DECLARATION OF INTERESTS}

The authors declare the following competing financial interest(s): Hai-Feng Li, Hui Kwun Nam, Pengfei Zhou, Shunping Ji, Kaitong Sun, Junchao Xia, Si Wu, and Yinghao Zhu have a 2021 China Invention Patent (patent number: CN114122509A) through the University of Macau based on this work: a solid-state electrolyte of ceramic oxide and its preparation method.

\section*{ACKNOWLEDGMENTS}

The authors acknowledge financial support from the Science and Technology Development Fund, Macao SAR (File Nos. 0051/2019/AFJ, 0007/2021/AGJ, 0090/2021/A2, and 0049/2021/AGJ), Guangdong Basic and Applied Basic Research Foundation (Guangdong-Dongguan Joint Fund No. 2020B1515120025), University of Macau (MYRG2020-00278-IAPME, MYRG2020-00187-IAPME, and EF030/IAPME-LHF/2021/GDSTIC), Guangdong-Hong Kong-Macao Joint Laboratory for Neutron Scattering Science and Technology (Grant No. 2019B121205003), and the National Natural Science Foundation of China (51871232).

\bibliographystyle{elsarticle-num-names}
\bibliography{NASICON1}


\end{document}